\documentclass[acmtog]{acmart}
\acmSubmissionID{xxxx}

\usepackage{booktabs} 

\usepackage[ruled]{algorithm2e} 
\usepackage{multirow}
\usepackage{array}    
\usepackage{booktabs} 
\usepackage{indentfirst}

\SetAlFnt{\small}
\SetAlCapFnt{\small}
\SetAlCapNameFnt{\small}
\SetAlCapHSkip{0pt}

\acmJournal{TOG}

\begin{document}
\title{From Geometry to Culture: An Iterative VLM Layout Framework for Placing Objects in Complex 3D Scene Contexts}

\begin{teaserfigure}
  \includegraphics[width=\textwidth]{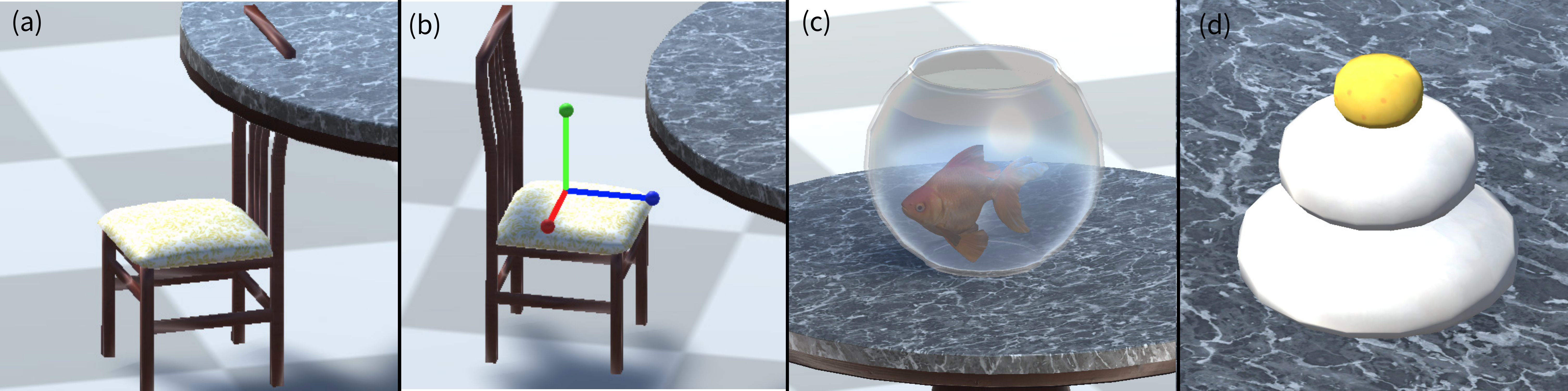} 
  \caption{(a) Result of placing a chair next to a table without using visual assitive cues (VACs).  
(b) Result when using VACs.  
(c) Result of placing a goldfish in a fishbowl, showing the system understands the context of placing it in water.  
(d) Result of stacking \text{Kagamimochi} (Japanese rice cakes), demonstrating the system’s ability to place objects based on local context.  }
  \label{fig:teaser}
\end{teaserfigure}

\author{Yuto Asano, Naruya Kondo, Tatsuki Fushimi, Yoichi Ochiai}


\begin{abstract}
3D layout tasks have traditionally concentrated on geometric constraints, but many practical applications demand richer contextual understanding that spans social interactions, cultural traditions, and usage conventions. Existing methods often rely on rule-based heuristics or narrowly trained learning models, making them difficult to generalize and frequently prone to orientation errors that break realism. To address these challenges, we define four escalating context levels, ranging from straightforward physical placement to complex cultural requirements such as religious customs and advanced social norms. We then propose a Vision-Language Model-based pipeline that inserts minimal visual cues for orientation guidance and employs iterative feedback to pinpoint, diagnose, and correct unnatural placements in an automated fashion. Each adjustment is revisited through the system’s verification process until it achieves a coherent result, thereby eliminating the need for extensive user oversight or manual parameter tuning. Our experiments across these four context levels reveal marked improvements in rotation accuracy, distance control, and overall layout plausibility compared with native VLM.  By reducing the dependence on pre-programmed constraints or prohibitively large training sets, our method enables fully automated scene composition for both everyday scenarios and specialized cultural tasks, moving toward a universally adaptable framework for 3D arrangement.
\end{abstract}

%
%


%
%


\maketitle
\section{Introduction}
\noindent
Placing objects is a highly context task. Humans do not merely fellow physical laws to decide where to place furniture in a room; we also consider affordances. For example, when we position a chair next to a table, we orient it toward the table. Similarly, when positioning a goalkeepr on a soccer field, we account for socially constructed rules of soccer by placing the goalkeeper in front of the goal opening. As the level of context increases, religious and cultural knowledge becomes necessary. For instance, Japanese shrines feature a pair of guardian statues called \textit{Komainu}, each with a distinct mouth shape; deciding which statue should be placed on the left demands, advanced contextual understanding. Although numerous systems have been proposed to generated indoor layouts based on user instructions or large training datasets, few have addressed layout design that requires such high-level context comprehension.

The placement systems themselves have existed since the dawn of computer graphics. Winograd introduced SHRDLU~\cite{WINOGRAD19721}, one of the earliest stems for manipulating objects in a three-dimensional environment via natural language. Rule-based approaches to object placement evolved through systems like Put~\cite{put} or WordsEye~\cite{wordseye}, which was already possible to construct a 3D scene from multiple natural-language sentences. Instead of relying on a dictionary of object relationships, probabilistic methods were later proposed to learn spatial relationships through 3D feature representations~\cite{scf,interactivefurniture,sceneseer,makeithome,AutomaticFurnitureRepresentation,InteractiveLearningOfSpatial,anInteractiveSystem}. With advances in neural networks such as CNNs~\cite{DeepConvolutionalPriors,fastAndFlexible}, VAEs~\cite{Grains}, GANs\cite{House-gan}, attention models~\cite{atiss,SceneGraphNet}, and diffusion models~\cite{TowardsTextGuided,DiffuScene,Diffusion-basedGeneration}-systems can now empirically infer plausible spatial relationships. Both dictionary-based and probabilistic approaches remain limited by the predefined set of objects they can handle; dictionary-based methods cannot cope with undefined objects or relationships, and probabilistic methods drop when encountering unfamiliar objects.

Methods leveraging large language models (LLMs) and vision-language models (VLMs) have proposed to solve this problem. LayoutGPT introduced a zero-shot approach for arranging 3D scenes. Aguina-Kang et al. enhanced LayoutGPT's accuracy and adopted the "open-universe" concept to successfully choose target objects from a large set. Though it needs training, LLplace enables interactive placement between user and system. Even with LLM/VLM-based methods, they do not exhibit sufficient generality for a wide range of layout tasks. LLplace requires prior training, while open-scene needs human-imposed constraints on object relationships. LayoutGPT remains limited to interior layout decisions. If we can improve VLM's placement capabilities, it may adapt to placement problems requiring advanced context. Furthermore, many VLM-based methods are not iterative optimizations but can specify all objects' parameters at once. Ideally, this could solve the placement problem in \textit{O(1)}, or \textit{O(N)} if parameters are set individually for each object. In current VLM-based placement systems, training and adding conditions manually limit these capabilities.

To address this issue, we propose a system that takes user instructions for object arrangement, determines appropriate positions, and place them accordingly. To apply a wide range of object placement tasks, it operates without pre-training. Our approach comprises two main components: a visual assistive cue (VAC) that helps the VLM identify objects, and a recursive routine that repeatedly checks for unnatural arrangements. To determine the most effective cues, we evaluated multiple tasks and found that markers indicating object axes and overhead views assisting scene comprehension significantly enhance placement accuracy. By leveraging VAC to detect layout anomalies, we can autonomously construct natural scenes without human involvement. In addition, we devised 12 arrangement tasks by introducing the concept of "context levels" to evaluate how well our model can handle diverse scenarios.
The main contributions of this work as are summarized as follows:
\begin{enumerate}
    \item We propose a desirable form of VAC to help VLMs more effectively detect discomfort in object arrangements and validate it across multiple tasks.
    \item We demonstrate that recursively having the VLMs judge recursively enables object arrangement from natural language instructions without pre-training.
    \item We devised multiple arrangement tasks for different context levels and evaluated our model.
    \item We confirmed that tasks such as people layout or object generation in outdoor can be tackled by non-procedural methods.
\end{enumerate}

\begin{figure*}[htbp]
  \centering
  \includegraphics[width=\textwidth]{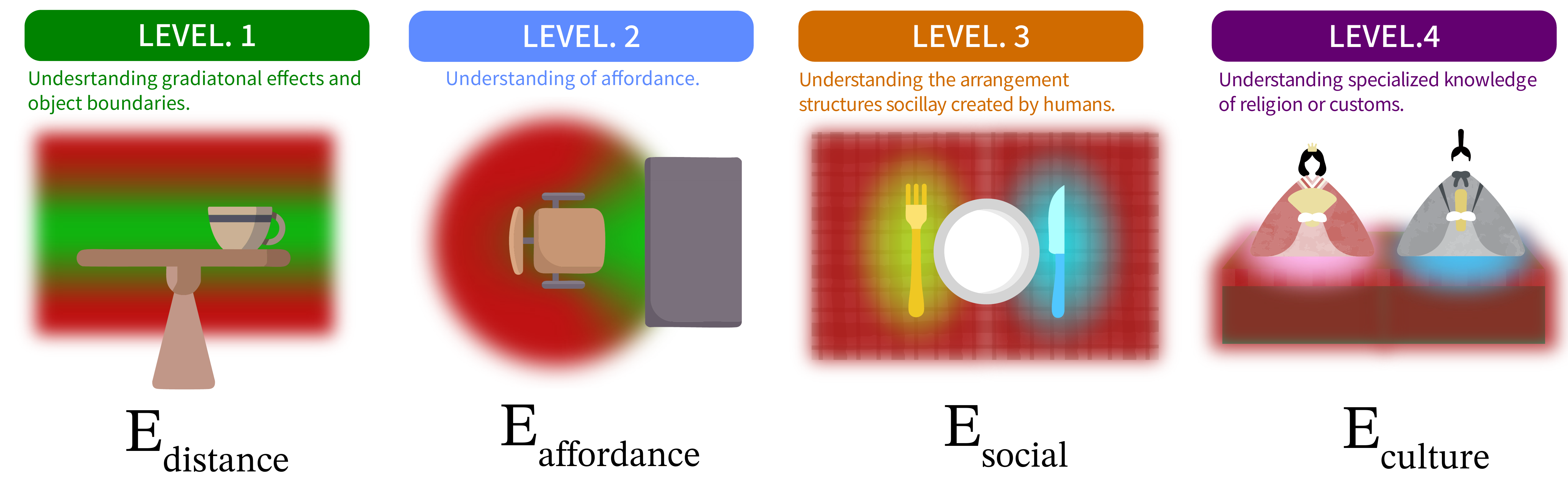}
  \caption{There are four context levels in arrangement tasks.  
Level 1 optimizes object parameters considering physical constraints.  
Level 2 adds affordance adjustments, such as orienting a chair toward a desk.  
Level 3 goes beyond affordances to social norms, for example placing a knife to the right of a plate with the blade facing inward.  
Level 4 includes religious and cultural constraints understood only by certain groups.  }
  \label{fig:def}
\end{figure*}

\section{Related Work}


\paragraph{Text to 3D-Scene Synthesis via Natural Language Processing}
Early approaches to text-driven 3D scene synthesis relied on natural language processing techniques. Put System \cite{clay1996put} and WordsEye \cite{DBLP:conf/siggraph/CoyneS01} pioneered this direction by mapping linguistic descriptions to predefined 3D scenes through manual rules and spatial relationships, but often resulting in unnatural scene descriptions. Some research has achieved naturalness by extracting relevance from data. \cite{DBLP:conf/iccv/ZitnickPV13} extract relationships between visual features and semantic meanings from a large number of image-text pairs. In \cite{DBLP:conf/emnlp/ChangSM14, InteractiveLearningOfSpatial}, their system learned probable spatial relationships and relative poses as probability distributions. Their follow-up work \cite{DBLP:conf/acl/ChangMSPM15} further improved this by introducing contrastive learning to capture the relationships between natural language words and 3D assets. \cite{DBLP:journals/corr/ChangESM17} extended this framework to enable interactive text-based scene editing, improving the user interface. However, these approaches required extensive preparation of training data and showed limited generalization beyond their training domains.

\paragraph{Neural Approaches to Scene Generation}
As highlighted in the survey \cite{DBLP:journals/corr/abs-2304-03188}, in the past few years have seen a surge in neural approaches to scene generation, particularly focusing on indoor environments. RoomDreamer \cite{DBLP:conf/mm/SongCXKTYY23} employs NeRF-based techniques to jointly generate layout, geometry, and textures, while LEGO-Net \cite{DBLP:conf/cvpr/WeiDPSP0G23} uses a diffusion-based approach to progressively refine initial placements into coherent arrangements. ATISS \cite{atiss} introduces an autoregressive transformer approach, predicting object placements one at a time while considering previous placements. Despite these impressive advances, these methods face a fundamental limitation: they rely heavily on available 3D scene datasets, which are primarily limited to indoor environments. Collecting ground-truth 3D scenes for open-domain scenarios remains extremely challenging, which has constrained the generalization capabilities of these approaches.

\paragraph{Large Pre-trained Models for Scene Generation}
Recent works have explored leveraging large language models (LLMs) for scene generation. LayoutGPT \cite{layoutgpt} SceneScript \cite{avetisyan2025scenescript}, and LLplace \cite{llplace} focus on generating scene descriptions that can be translated into 3D arrangements.
Scene Language \cite{zhang2024scenelanguage} combines the description of scenes with the script generation and integration of visual embedding, then realizes the natural arrangement task of a wide range of objects, not limited to furniture arrangement.
Holodeck \cite{yang2024holodeck} leverages a pre-trained LLM to handle asset selection, placement, and design adjustments based on text instructions, enabling fully automated 3D environment generation and editing. The LLM enumerates the constraints on spatial relationships among objects, and a depth-first search starting with the larger objects then optimizes their arrangement to satisfy those constraints. However, unlike purely neural approaches, each object's position and rotation are not directly computed as continuous numerical values; instead, they are determined from discrete candidates for spatial relationship labels, positions, and rotations.
3D-GPT \cite{DBLP:journals/corr/abs-2310-12945} has incorporated the LLM agent into Blender to automate the generation and modeling of natural scenes, while LLMR \cite{De_La_Torre_2024} focuses on generating Unity scene construction. These techniques also employ LLM agents that generate python and C\# scripts, and realize more automatic and detailed object control, such as referencing the state in various scenes or executing repetitive process.
SceneCraft \cite{hu2024scenecraft} further utilizes pre-trained Vision language models (VLM) to evaluate scene placement, and iteratively optimizes scenes by rendering and visually providing feedback on the generated scenes. It has succeeded in generating a wide range of natural scenes, from ``stacked boxes'' to ``airport terminals''. However, on closer examination, one finds that relatively few scenes require precise relative positioning or object rotations; perhaps most tasks can be completed with relatively little difficulty, such as arranging chairs at equal intervals or randomly distributing people. In this study, we propose a series of placement tasks categorized into multiple levels, with the aim of accurately evaluating scene generation methods by varying the complexity of spatial arrangement. We also employ simplified versions of SceneCraft that combine LLMs and VLMs as baselines, so that we can isolate the method’s complexity from the tasks’ inherent difficulty.

A critical consideration when employing large pre-trained models is their limited ability to extract specialized information, particularly for precise object placement. This limitation arises because most large models are trained on open-domain data. It becomes especially evident in in-the-loop optimization processes, in which iterative feedback from vision models is used to adjust object placement, because accurate recognition of object rotation is crucial. Although several works \cite{zhou20233d, wang2021gdr, hara2017designing} have acknowledged this challenge and released rotation-annotated datasets \cite{xiang2014beyond}, these datasets remain significantly smaller than general-purpose vision datasets, both in terms of category coverage and sample size. Consequently, when designing systems, it is key not to rely too heavily on the rotation recognition capabilities of these vision models. Addressing this issue at the system architecture level can help ensure robust performance in tasks that require rotational understanding.

\begin{figure*}[htbp]
  \centering
  \includegraphics[width=\textwidth]{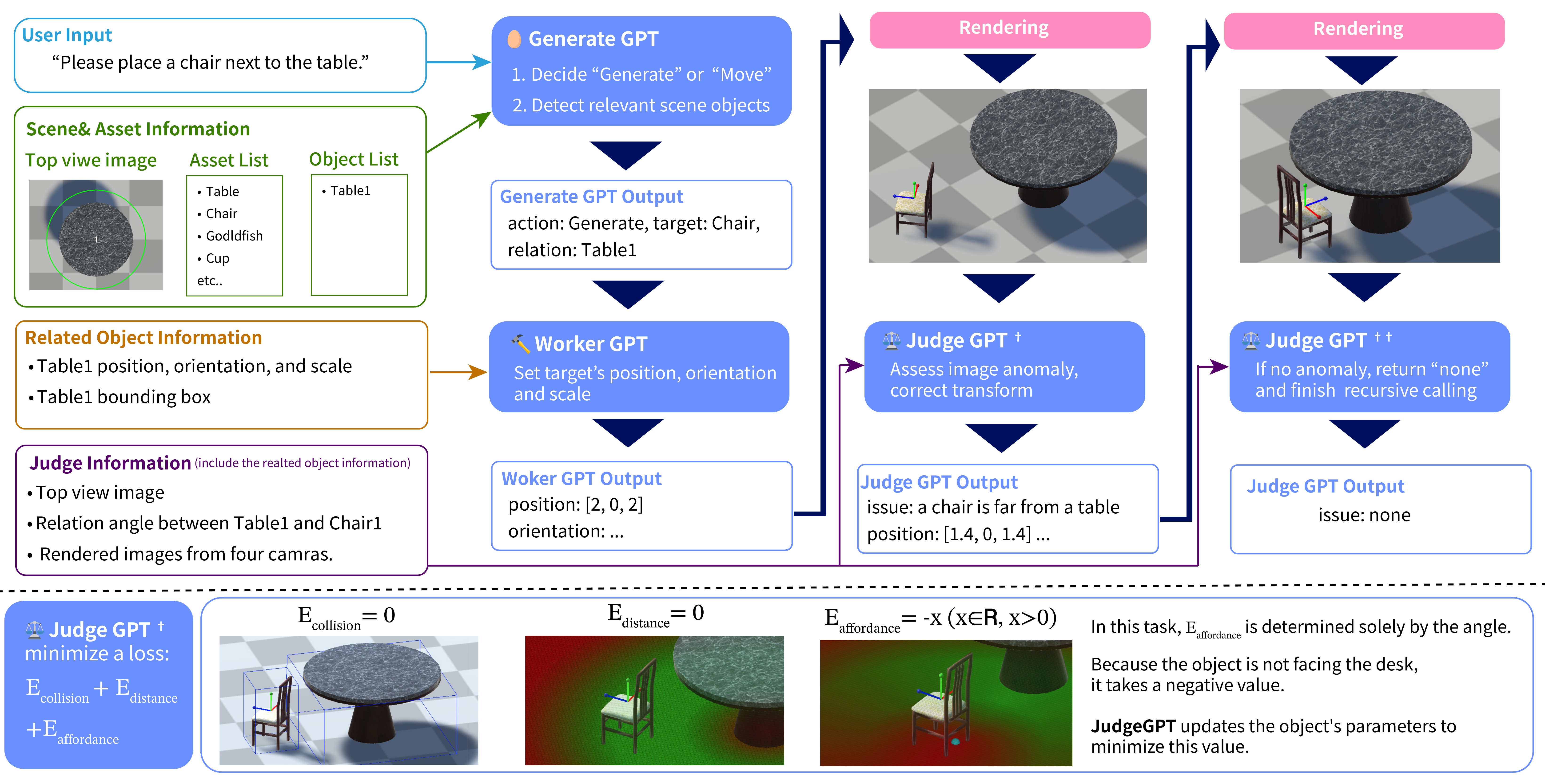}
  \caption{Our system architecture. Thick arrows indicate the execution sequence, while thin arrows the supplementary information attached to each GPT request. The judge information includes the related object information. Although \textit{JudgeGPT} is shown being called twice in this figure, it continues to run as long as inconsistencies in the scene arrangement are detected. \textit{JudgeGPT} virtually updates the object's parameters to minimize the sum of three losses ($E_{collision\_total}+E_{distance}+E_{affordance}$). As the context level increases, additional error terms appear, requiring further optimization.}
  \label{fig:oveview}
\end{figure*}

\section{Problem statement}
\noindent
The task of object placement is to determine each object’s position \( \mathbf{p}_i \in \mathbb{R}^3 \), rotation \( \mathbf{r}_i \in SO(3) \), and scale \( \mathbf{s}_i \). In conventional methods, these parameters are typically determined by defining cost functions primarily related to collision avoidance or placement stability and minimizing them. For example, Merrell et al. define total collision loss optimization the Minkowski sum~\cite{interactivefurniture} as follows:
\[\min_{\{\mathbf{p}_i,\mathbf{r}_i,\mathbf{s}_i\}} \; 
E_{\text{collision}} \;=\; \sum_{(i,j)} A\bigl(J_{o_i}\,\cap\,J_{o_j}\bigr)\;
\]
where \( J_{o_i} \) is the Minkowski sum of the 2D ground-plane projection of object \( o_i \) with its recommended clearance shape, \( A(\cdot) \) measures the overlap area. Strictly speaking, it includes the region around the object called “clearance,” but setting its width to zero makes it equivalent to collision detection. However, such methods are specialized to geometric constraints and struggle to handle complex contexts such as cultural, social, or usage-based factors. While heuristic approaches (e.g., placing plates on a table because they are often found there) are sometimes adopted, systematic research to mathematically formalize these considerations is rare.

To address this problem, we propose a method for incorporating context—classified into Levels 1 to 4—into the optimization problem as additional costs or constraints as shown in Fig.~\ref{fig:def}. Level 1 ensures collision avoidance and suitable relative positioning, formulated as:



\begin{align}
\min_{\{\mathbf{p}_i, \mathbf{r}_i, \mathbf{s}_i\}} \; \Bigl(
    \sum_{(i,j)} E_{\text{collision}} 
    + E_{\text{distance}}
\Bigr), \quad \quad \quad \nonumber\\[6pt]
\text{where} \quad 
E_{\text{distance}} \;=\; 
\lambda_{\mathrm{dist}} \sum_{(i,j)}
\|\mathbf{p}_i - \mathbf{p}_j - \mathbf{d}_{ij}^*\|^2. \nonumber
\end{align}


 \( \lambda_{\mathrm{dist}} \) is the weight for the distance penalty. The vector $ \mathbf{d}_{ij}^* $ represents a relative position that incorporates physical constraints. For example, when placing a cup on a table, its horizontal placement may be anywhere on the table, but the vertical position must rest on the table surface. Level 2 requires the ability to position objects by considering affordances between them, as humans do. For example, beyond merely placing a chair next to a table, it also requires orienting the chair toward the table. We define
\[
E_{\text{affordance}} = \sum_{(i,j)}\max\Bigl(0,\; \alpha_{\text{affordance}} \cdot d_{\text{affordance}}\Bigr),
\]
where \(d_{\text{affordance}}\) $\in \mathbb{R}$ measures how well an object aligns with a given affordance, and \(\alpha_{\text{affordance}}\) is a weight for any misalignment penalty. The overall objective is then
\[
\min_{\{\mathbf{p}_i, \mathbf{r}_i, \mathbf{s}_i\}} 
\Bigl( 
  E_{\text{collision}} +  E_{\text{distance}}
  + E_{\text{affordance}}
\Bigr).
\]
At Level 3, we further evaluate whether objects maintain social relationships. For example, this involves correctly placing a soccer goal and goalkeeper, or arranging multiple objects in a classroom. Following the Level 2 treatment of affordances, we introduce an additional error term $E_{\text{social}}$ for the deviation from proper social positioning and minimize it as:
\[
\min_{\{\mathbf{p}_i, \mathbf{r}_i, \mathbf{s}_i\}} 
\Bigl( 
  E_{\text{collision}} +  E_{\text{distance}}
  + E_{\text{affordance}} +E_{\text{social}}
\Bigr).
\]
At Level 4, we require a culture fit that accounts for religious or traditional contexts. For instance, this level evaluates whether Japanese guardian dogs (Komainu) can be placed appropriately at a Shinto shrine. Similarly, this task add the term $E_{\text{culture}}$ and minimize it.

\section{System Overview}
\noindent
Our system aims to generate coherent scenes based on user-provided placement instructions expressed in natural language. To achieve this, we employ a recursive GPT request process combined with a visual assistive cue (VAC; see Section~\ref{sec:vac}). Fig.~\ref{fig:oveview} presents our framework overview.

When the user inputs an object arrangement direction, the system creates text-based lists of generalizable assets and existing objects in the scene. It also obtains a top view parallel-projection image. These data, together with instruction, are passed to the \textit{GenerateGPT} request, which determine whether the instruction requires creating a new object or moving an existing one. If a new object is needed, the system selects it from the asset list; if an existing object must be moved, it is chosen from the object list. This object is called target object. In both cases, any relevant object involved in the spatial relationship (for instance, a table when placing a chair next to it) is also identified.

Once the target object has been determined, the system gets the relevant object's position, orientation, scale and bounding box information. It then invokes the \textit{WorkerGPT} request to infer the target object's transformation parameters and updates these values in the scene. Afterward, the system obtains an updated top view image, calculates relative angles to each related object, and rendered the target object from four surrounding viewpoints. These results are sent to the \textit{JudgeGPT} module to evaluate if the scene appears natural. If any unnaturalness is detected, the system corrects the target object's parameters and repeats the evaluation. When the \textit{JudgeGPT} finds no further inconsistencies, the placement operation is deemed complete. We used the same model for \textit{GenerateGPT}, \textit{WorkerGPT}, and \textit{JudgeGPT}. Specifically, this study employed OpenAI GPT-4o and o1. The entire system was built in Unity and C\#.

\section{Visual assistive cues}~\label{sec:vac}
\noindent
In order to enable our system to arrange objects in natural way, it must be capable of detecting when its own arrangements appear unnatural. In the early stage of this research, we tested whether a VLM could detect anomalies from a single rendered image of a scene without any additional information. Our results showed that, compared to humans, the VLM's ability to identify such anomalies was markedly inferior. For instance, when placing a chair next to a desk, a human would typically ensure that the chair is both close to and facing the desk wo that a seated person could reach objects on the table. In contrast, the VLM often concluded that arrangements were acceptable even if the chair was far from the table or oriented incorrectly (e.g., as in Fig.1(a)). This inability to detect unnatural placements poses a challenge in zero-shot scene arrangement. If the VLM could detect anomalies independent of object categories and shapes, it would open the way to a zero-shot system capable of creating more natural layouts.

To address this issue, we propose visual assistive cues (VAC), which provide supplemental information -either as text or images-in addition to the user inputs and the essential components of the scene. We examined various types of VAC to determine which were most effective at improving detection accuracy, selecting cues that yielded the most substantial gains. There are of course countless possible forms of VAC, and we admit we only tested a small subset in this study.

\begin{figure}
  \centering
  \includegraphics[width=0.45\textwidth]{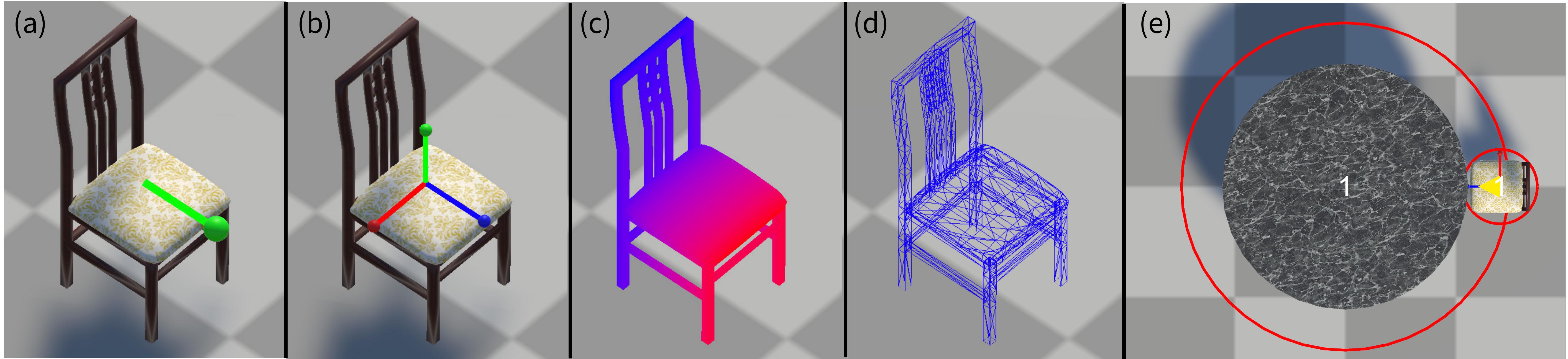}
  \caption{VAC candidates delivered as images. The front maker (a) renders only the front view, while the other version (b) includes three axes. (c) applies a front shader, and (d) shows the wireframe. (e) depicts Clearance circles and indices for both the table and chair.}
  \label{fig: vacs}
\end{figure}

\subsection{VAC Candidates}
\noindent
We devised several types of VAC and fig.~\ref{fig: vacs} summarizes the VACs provided as images.

\textit{Front marker.} VLM has limited ability to detect object orientation in images, often confusing the front and back of chairs in a scene. We tried to attach markers to each object to indicate its front, ensuring they appear upon rendering. Two marker types were prepared: one denoting only the front, and another set comprising markers for the front, top, and right. The latter set is inspired by the transform tools used in Unity and other CG software for positioning objects.

\textit{Front shader.} Instead of using markers to distinguish an object's front and back, we attempted to clarify its front-back orientation by gradually coloring the front portion in red and the back potion in blue.

\textit{Wireframe} To remove the influence of textures and infer orientation purely from the object's shape, we also evaluated only wireframe rendering.

\textit{Relation Angle.} Even if a front marker identifies the front face of an object, multiple steps are still needed to determine the exact orientation required for it to face the related object. To reduce the necessity for LLM-based inference, we provide textural information about the angle toward the object's center and how many degrees the current object's front is offset. This offset is calculated as the difference between (1) the rotation angle of the vector connecting the centers of bounding boxes of the related object and target objects, and (2) the orientation angle of the target object.

\textit{Bounding box.} All object parameters can be described solely by position, orientation, and scale. If a VLM could fully infer overlaps and distances among objects from an rendered image, no additional distance information would be necessary. However, because a VLM alone struggled to discern detailed spatial relationships, we attempted to provide bounding box information for related objects through textual input. Such bounding box information is also employed by other VLM-based layouts systems: in LLplace~\cite{llplace}, it is part of the input/output data, and in LayoutGPT~\cite{layoutgpt}, objects are arranged by specifying their bounding boxes.

\textit{Clearance circle and index.} Humans naturally avoid placing objects ins overly close proximity and maintain a suitable distance between them. To help a VLM determine whether sufficient space exists between objects, we introduce a "clearance circle"- a visual marker inspired by classical arrangement systems~\cite{interactivefurniture}. Specifically, we draw a circle at a fixed offset from each object (one-third of its bounding box radius in our research). Whenever two circles collide, they change color from green to red. In addition, when multiple objects of the same type appear in a single scene, the VLM has difficulty distinguishing which object corresponds to which label. To address this, we add an index at the center of each clearance circle. Both the clearance circles and the indices are rendered in top view.

We empirically found that the following combination of markers is effective: bounding box, relation angle, front marker (triple), clearance circle, and index as VAC. Differences among the markers are presented in the appendix.



\begin{figure*}[htbp]
  \centering
  \includegraphics[width=\textwidth]{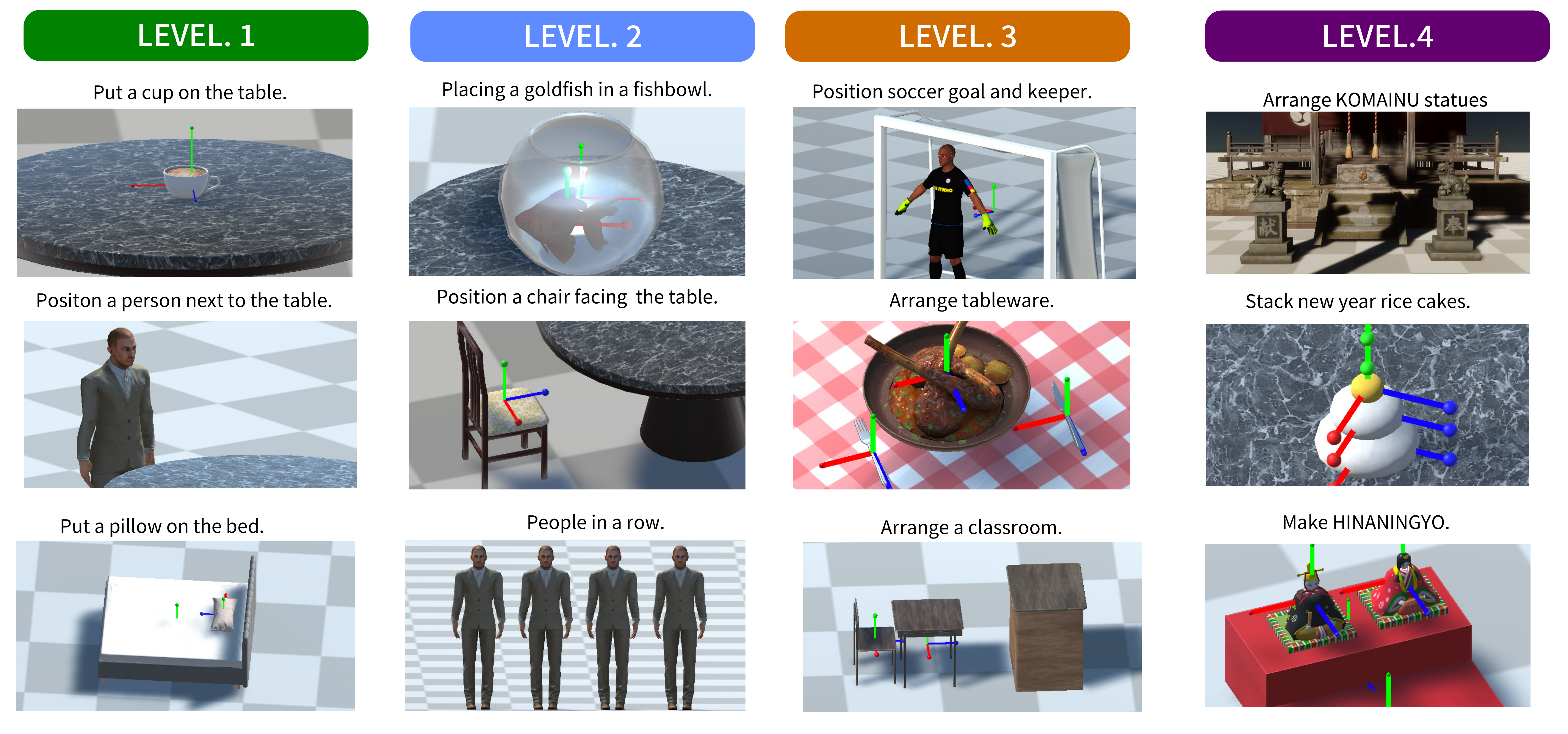}
  \caption{12 placement tasks with different context levels. As the level increases, social awareness and specialized knowledge become necessary. The number of objects to be placed ranges from one to four.}
  \label{fig: contextLevel}
\end{figure*}

\begin{table*}[h]
\centering
\begin{tabular}{|l||c|c|c||c|c|c||c|c|c||c|c|c|}
\hline
 & task1 & task2 & task3 
 & task1 & task2 & task3 
 & task1 & task2 & task3 
 & task1 & task2 & task3 \\
\hline
accuracy [\%] &100 &100 &40 &90 &90 &80 &20 &0 &60 &0 &100 &0 \\
\hline
speed [s] &69.4 &136.3 &125.9 &172 &154.6 &350.6 &151 &363.6 &426.9 &496.2 &327 &283.3 \\
\hline
loops &1 &2.1 &1.6 &2.1 &1.8 &5.4 &2 &5.4 &5.1 &5.6 &3.9 &4.2 \\
\hline
\end{tabular}
\caption{Our systems results on 12 tasks, each with 10 trials. The reported values are averages across these trials. Accuracy is the mean of binary measure (1 for success, 0 for failure). Speed is measured from the moment the natural language command is given until the arrangement is complete. Loops indicates the number of \textit{JudgeGPT} called.}
\label{tab:result_table}
\end{table*}

\section{Results}
\noindent
We evaluated our system using 12 tasks. These tasks are structured as three tasks, each with four context levels. Every context level always includes at least one human or doll. We also designed them to feature varied object scales and outdoor scenes. All task images are provided in Fig.~\ref{fig: contextLevel}, and the appendix contains detailed explanations and success criteria. All evaluation results are shown in Table.~\ref{tab:result_table}. Each entry represents the average of 10 trials.  

\begin{figure*}
  \centering
  \includegraphics[width=\textwidth]{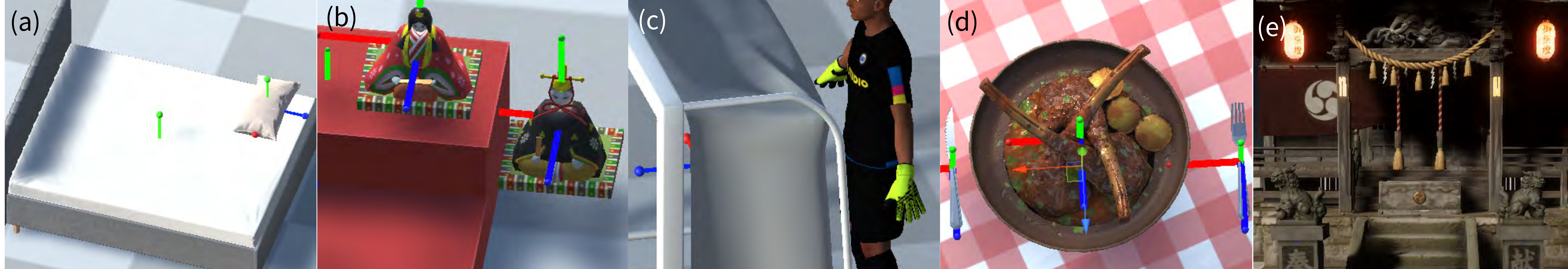}
  \caption{Major failures of this system:  
(a) Placing the pillow higher than the mattress  
(b) \textit{Hina} dolls floating in midair  
(c) Keeper and goal reversed  
(d) Knife and fork swapped left to right  
(e) \textit{Komainu} statues swapped left to right  }
  \label{fig: add}
\end{figure*}

 At context level 1, all tasks except "place a pillow on the bed" exceeded 90\% success.  "Place a cup on the table" and the horizontal "place a person next to the table" tasks recorded similar accuracy, suggesting no major difference between vertical and horizontal accuracy. However, placing a pillow on the bed succeeded only about 40\% of the time (Fig.~\ref{fig: add}~(a)). This suggests that the bed's bounding box is calculated at twice the mattress height, causing the model to misinterpret this higher position as the mattress. Although the pillow is visibly floating, the text-based bounding box input appears to override this visual cue.
 
 At context level 2, all tasks achieved over 80\% accuracy. Since no inconsistencies like the pillow-placement task at level 1 arose, task execution remained stable. Considering that existing methods struggle with tasks at this level, this model appears to have sufficient context-understanding capability.  

At level 3, placement accuracy drops significantly. The highest accuracy was achieved in Task 3, placing three objects in a classroom, with a success rate of 60\%. Common errors included cases where the teacher's desk, which should face the student's desk, was reversed and when student chairs intruded into the teacher's desk. The system's accuracy decreases as the number of objects increases of their relative positions are not evenly arranged (e.g., not a regular intervals or angels). All attempts at placing the knife and fork failed in the same way~(Fig.~\ref{fig: add}~(d)). Although the system correctly parsed the instruction "knife on the right and fork on the left of the plate", it treats the negative coordinate direction as the default front for both utensils. As a result, placing the knife on the right actually puts it on the left. Even though the knife appears on the left visually, it is registered with a positive coordinate value relative to the plate in the text data, suggesting that the textual information takes precedence. A similar issue is believed to have occurred with the soccer goalkeeper and goal (Fig.~\ref{fig: add}~(c)). In addition, the goal's transparent material may have effected the accuracy of visual detection.

Tasks at level 4 were nearly impossible to execute. In the task of placing \textit{Komainu} statues~(Fig.~\ref{fig: add}~(e)), the instruction themselves frequently got their left-right positioning wrong. This suggest that as context level increases, the VLM may not possess the required knowledge. In the \textit{Hina} dolls placement task, about half of the instructions were correct(Fig.~\ref{fig: add}~(b)). However, in some cases, the first doll was placed in the center of the tier or floating in the air, deciding the space for the second doll was difficult. This indicates that in scenes with few references, it is difficult to infer specific coordinates from instructions like "on the right side of the upper tier." The \textit{Kagami-mochi} placement task never failed. This is similar to placing a cup on a table, indicating that, beyond the complexity of the context, the physical arrangement heavily influences task success.

\section{discussion}
\subsection{Our System Limitation}
\noindent
Our system excels at placing objects with uniform lengths or angles and at processing them according to bounding boxes. However, as the number and complexity of object relationships increase, it fails to capture all relationships, resulting in a rapid drop in accuracy. We attribute this to the VLM's attention mechanism, which struggles with diverse positional relationships. When generating a scene step by step based on user input-rather than creating it one shot via the VLM-this limitation becomes a bottleneck. In the future, we need to add a system that compensates for this shortcoming.

Tasks like placing pillows within a system-computed bounding box exemplify a typical failure~(Fig.~\ref{fig: add}~(a)). Removing the bounding box cue could fix this, but altering cues only for certain tasks introduces human biases, so some extension-such as letting the LLM decide-is needed. Also, even if the visual arrangement appears problematic, textual inferences about physical contact (via bounding boxes) can dominate the final output. This indicates the system has not yet fully developed the ability to integrate multiple modalities and vive the entire scene holistically.

When the front of an object is misaligned, proper placement becomes difficult~(Fig.~\ref{fig: add}~(c,d)). As noted in the results, while the system could generate language instructions for placing a knife and fork, it failed to visually recognize the ``right side of the plate.'' Because the plane on which objects are placed does not always align with the left-handed coordinates system, a plane transformation is necessary to solve this problem.

\subsection{Layout tasks at different context levels}
\noindent
We introduced the concept of ``context level'' for placement tasks and evaluated our system. Even though the number of objects did not vary greatly, placement accuracy decreased as the context level rose, suggesting our level definitions were appropriate. This concept highlights that current VLMs cannot handle high-level placement tasks, and future systems must overcome this. However, large accuracy gaps emerged within the same level due to geometric constraints, such as placing a pillow on a bed or correctly aligning object fronts. Developing a better evaluation matrix that integrates geometric conditions with context levels remains a key challenge.

The context levels concept is not limited to object arrangement; it can be applied broadly to CG tasks such as 3D object generation, 2D rendering, font selection, or motion generation. Given that LLMs and VLMs are now employed for an array of tasks, incorporating context levels into these domains seems beneficial. In animated films, viewers often find certain on-screen fonts jarring or question motions that would be unusual for someone accustomed to that culture. While LLms and VLMs are now being introduced as CG solutions, we must go not only solving existing problems in a zero-shot, but also consider how to utilized these models to design and solve problems with high contextual demands.

\subsection{Unexplainably nature of VLMs}
\noindent
VLMs remain black boxes, making it impossible to fully explain or directly improve fluctuations in our system's accuracy. This is highly inconvenient from a academic evaluation perspective. Yet VLM-based methods are currently the only ones capable of tackling high-context-level arrangements tasks, so we adopted a VLM-based system despite its black-box nature. One way to mitigate the black-box nature of VLMs is to visualize the prompts and image embeddings used for arrangement, but there is currently no publicly available VLM that provides such image embeddings, leaving this verification as a future challenge.

When neural networks were first introduced into computer graphics, their black-box nature also drew criticism. However, as their performance improved and they began reliably solving problems, they became widely accepted. With LLMs and VLMs advancing rapidly, we expect them to tackle an ever-growing range of CG tasks. Our study represents an early step in this progression, demonstrating that by incorporating multiple levels of context into these tasks, we can thoroughly evaluate  current VLM performance.

\subsection{Layout speed}
Our system requires at least two minutes to place a single object, or about three minutes if angle adjustments are involved-speeds too slow to ensure sufficient interactivity. However, when solving high context level problems that can tolerated longer runtimes, the system excels. For instance, it could be paired with procedural generation to synthesize mid-scale scenes over extended periods. 

In contrast, humans can instantly envision an object's ideal placement (though minor adjustments may be needed), meaning they do not need recursive placement process. Ideally, batch-editing all objects would take \textit{O(1)} time, while editing them individually could take \textit{O(N)}. As VLM performance improves, the system could reduce the number of \textit{JudgeGPT} calls and achieve layout generation speeds that support a more interactive workflow.


\bibliographystyle{ACM-Reference-Format}
\bibliography{sample-bibliography}

\appendix
\section{Layout evaluation tasks detail}
\noindent
Explain the details of the layout tasks corresponding to the context levels.

When arranging furniture or other objects within a given space, humans do not rely solely on physical constraints but also take into account various the chair toward the desk. Similarly, when positioning a goalkeeper on a soccer field, the goalkeeper is placed in front of the goal net in accordance with general soccer rules. As the contextual level increases, deeper religious and cultural knowledge may become necessary. For instance, at a Japanese Shinto shrine, there is a pair of guardian statues called \textit{"komainu"}, each with a different mouth shape, Correctly determining which statue should be placed on the left side requires a high level of contextual understanding.
Although many systems have been proposed to automatically generate indoor layouts based on user instructions and large-scale training datasets, only a few have tackled layout design that demands this higher level of contextual reasoning.
In this work, we conduct a series of experiments designed to evaluate object placement within a space, considering not only physical constraints but also social rules and cultural backgrounds. We define four levels of increasing complexity to systematically assess the performance of layout systems.

\subsection{Context Level.1}
\noindent
The task, Level.1,  focus on fundamental physical placement.

\textit{"Put a cup on the table."}  
A system has to place a cup on an existing table. The cup must fit on the table's surface in the horizontal plane, and its vertical position must remain within a predefined threshold.

\textit{"Put a person next to the desk."} 
A system has to place a person within a certain distance from the table, without any restriction on orientation.

\textit{"Put a pillow on the bed."} 
A system has to place a pillow on an existing bed. Any horizontal position on the bed is acceptable, and the vertical position must remain within the threshold.

\subsection{Context Level.2}
\noindent
Level 2 introduces slightly more advanced  physical conditions.

\textit{"Place a goldfish in a fishbowl."}
A system has to place a goldfish inside a goldfish bowl, which itself can be placed anywhere on a table within the vertical threshold, The goldfish must be located within the water portion of the bowl.

\textit{"Place a chair next to the desk."}
A system has to position a chair within a certain distance from the desk and ensure that it faces toward the desk.

\textit{"Line up a group of people."}
A system has to arrange multiple people so they all face the same direction, and the distance between adjacent individuals stays within a specified threshold.

\subsection{Context Level.3}
\noindent
Level 3 tasks requires consideration of social rules and customary practices.

\textit{"Place a soccer goal and a goalkeeper on the field."}
A system has to place the goalkeeper in front of the soccer goal, ensuring the goalkeeper faces away from the goal(i.e., toward the field).

\textit{"Arrange tableware according to table manners."}
A system has to place the main dish in the center, with the knife on the right side and the fork on the left side.

\textit{"Arrange a classroom layout"}
A system has to place the teacher's desk, and student desks, so that they face each other, ensuring their horizontal distances are not drastically misaligned.

\subsection{Context Level.4}
\noindent
Level 4 tasks involves tasks that cannot be correctly executed without an understanding of traditional events or religious customs.

\textit{"Place a pair of KOMAINU statues at a Shinto shrine."}
A system has to place two lion-dog guardian statues in the correct position. These two statues are known as \textit{"Komainu"} in Japanese Shinto shrines such that the statue featuring an open mouth ("a") on the right side and the statue with a closed mouth ("un") on the left side.

\textit{"Assemble a KAGAMI-MOCHI."}
A system has to stack a large rice cake, then a smaller rice cake on top of it, and finally place a mandarin orange at the very top. This arrangement is called \textit{"Kagami-Mochi"} in Japan, a traditional New Year's decoration composed of two layers of \textit{"Mochi"} (rice cake) and crowned with a citrus fruit.

\textit{"Set up a HINA-MATSURI display."}
A system has to place \textit{"Odairi-sama"} (the male doll) and \textit{"Ohina-sama"} (the female doll), which together represent the couple in Japan's traditional event(\textit{"Hina-Matsuri"}) display, in a symmetrical arrangement on the higher tier of a two-tier platform.

\section{VACs evaluation}\label{sec:vaceval}
\noindent
In order to evaluate the accuracy of each VAC, we executed the task \textit{``Place a chair next to the desk.''} ten times for every combination of VAC. For each execution, we recorded (i) the task success rate, (ii) the time required to complete the arrangement, and (iii) the number of loops used by the LLM to perform the arrangement, following the same procedure as in our main experiment. While we used openAI's o1 model in the main experiment, we employed the gpt-4o model here to more clearly record the visual differences among VACs by eliminating intermediate inference steps. The results are presented in Table~\ref{tab:experiment_results}.

From the results, we observe that the success rates for tasks 3, 4, 6, and 7 each achieved 100\%. Although the single-marker approach completes tasks in a shorter time than the triple-marker approach, the latter increases the likelihood of successful recognition by the camera. Therefore, our implementation adopts the triple-marker approach.

When using a wireframe representation, the \textit{JudgeGPT} system deemed the object’s appearance unnatural, causing measurements to continue indefinitely and preventing proper record keeping. This situation—being judged as unnatural—rarely occurred with the shader setting, and when it did, it was more likely to be correctly recognized in the subsequent judgment. Thus, it merely added a small increase to the overall measurement time. Furthermore, our empirical observation suggests that top-view images can improve accuracy in tasks requiring precise alignment. Hence, in our final setup, we combined the bounding box, related angle, triple markers, and top-view elements to construct the chosen VAC.
\begin{table}[htbp]
  \centering
  \begin{tabular}{lccc}
    \hline
    method & accuracy & time & judge counts \\
    \hline
    1. none           & 0   & 59.5 & 4.1 \\
    2. BB              & 10  & 77.3 & 5.6 \\
    3. TRIPLE+RA+BB       & 100 & 71.6 & 5.1 \\
    4. SINGLE+RA+BB      & 100 & 46.9 & 2.8 \\
    5. WF+RA+BB       & -   & -    & -   \\
    6. SD+RA+BB       & 100 & 81.1 & 6.1 \\
    7. TRIPLE+RA+BB+TOP   & 100 & 99.1 & 5.6 \\
    \hline
  \end{tabular}
  \caption{Results of VACs evaluations.}
  \label{tab:experiment_results}
\end{table}

\end{document}